\documentclass[aps,pra,reprint,a4paper,superscriptaddress,
  balancelastpage]{revtex4-1}

\pdfoutput=1

\usepackage[UKenglish]{babel}
\usepackage[UKenglish,cleanlook]{isodate}
\usepackage{amsmath}
\usepackage{amssymb}
\usepackage[retainorgcmds]{IEEEtrantools}
\usepackage{dsfont}
\usepackage{paralist}
\usepackage{tikz}
\usepackage[colorlinks,linkcolor=blue,citecolor=blue,urlcolor=blue]{hyperref}
\usepackage{quant_defs}

\def\bsmf{Bull.\ Soc.\ Math.\ Fr.}%
\def\ieeetit{IEEE Trans.\ Inf.\ Th.}%
\def\josab{J.\ Opt.\ Soc.\ Am.\ B}%
\def\nature{Nature}%
\def\natcomm{Nat.\ Commun.}%
\def\njp{New J.\ Phys.}%
\def\pra{Phys.\ Rev.\ A}%
\def\prl{Phys.\ Rev.\ Lett}%
\def\prx{Phys.\ Rev.\ X}%
\def\prsa{Proc.\ R.\ Soc.\ A}%
\usetikzlibrary{shapes.misc,shapes.symbols,shapes.geometric,arrows,
  plotmarks,decorations.pathmorphing}
\tikzset{gridlines/.style={very thin,color=gray!50},
  every picture/.append style={scale=0.7}}

\usetikzlibrary{shapes.misc,shapes.symbols,shapes.geometric,arrows,
  positioning,plotmarks,decorations.pathmorphing}
\pgfdeclarelayer{background}
\pgfsetlayers{background,main}
\tikzset{gridlines/.style={very thin,color=gray!50},
  leftshift/.style={xshift=-5mm},
  rightshift/.style={xshift=5mm},
  upshift/.style={yshift=5mm},
  downshift/.style={yshift=-5mm},
  source/.style={rectangle,thick,draw=blue!50,fill=blue!20,thick,
    inner sep=0pt,minimum height=10mm,minimum width=10mm},
  detector/.style={rectangle,thick,draw=blue!50,fill=blue!20,thick,
    inner sep=0pt,minimum height=10mm,minimum width=10mm},
  interact/.style={rounded rectangle,draw=black,thick,
    inner sep=0pt,minimum height=10mm,minimum width=14mm},
  post/.style={->,shorten >=1pt,>=stealth',very thick},
  every picture/.append style={scale=1}}

\definecolor{darkgreen}{rgb}{0,0.5,0}

\newcommand{\rA}{\mathrm{A}}
\newcommand{\rB}{\mathrm{B}}

\newcommand{\rE}{\mathrm{E}}

\newcommand{\rABE}{\mathrm{ABE}}

\newcommand{\rS}{\mathrm{S}}
\newcommand{\rM}{\mathrm{M}}
\newcommand{\rw}{\mathrm{w}}
\newcommand{\rx}{\mathrm{x}}
\newcommand{\ry}{\mathrm{y}}
\newcommand{\rz}{\mathrm{z}}
\newcommand{\rmin}{\mathrm{min}}
\newcommand{\hilb}{\mathcal{H}}

\newcommand{\ee}{\mathrm{e}}
\newcommand{\ii}{\mathrm{i}}

\newcommand{\0}[0]{^{\vphantom{\prime}}}
\newcommand{\1}[0]{^{\prime}}

\begin{document}

\title{Secrecy in prepare-and-measure CHSH tests with a qubit bound}
\author{Erik Woodhead}
\date{7 October 2015}
\email{Erik.Woodhead@icfo.es}
\affiliation{ICFO -- Institut de Ci\`e{}ncies Fot\`o{}niques,
  av.\ Carl Friedrich Gauss 3, 08860 Castelldefels (Barcelona), Spain}
\author{Stefano Pironio}
\affiliation{Laboratoire d'Information Quantique, CP~224, Universit\'e{}
  libre de Bruxelles (ULB), 1050 Bruxelles, Belgium}

\begin{abstract}
  The security of device-independent (DI) quantum key distribution (QKD)
  protocols relies on the violation of Bell inequalities. As such, their
  security can be established based on minimal assumptions about the devices,
  but their implementation necessarily requires the distribution of entangled
  states. In a setting with fully trusted devices, any entanglement-based
  protocol is essentially equivalent to a corresponding prepare-and-measure
  protocol. This correspondence, however, is not generally valid in the DI
  setting unless one makes extra assumptions about the devices. Here we prove
  that a known tight lower bound on the min entropy in terms of the CHSH Bell
  correlator, which has featured in a number of entanglement-based DI QKD
  security proofs, also holds in a prepare-and-measure setting, subject only
  to the assumption that the source is limited to a two-dimensional Hilbert
  space.
\end{abstract}


\maketitle


The security of quantum key distribution (QKD) rests on tradeoffs inherent to
quantum physics, such as the impossibility of state cloning, the
measurement-disturbance tradeoff, or the monogamy of entanglement. Similarly,
the security of device-independent (DI) QKD
\cite{ref:e1991,ref:my1998,ref:ab2007}, which can be established with minimal
assumptions about the internal functioning of the devices, is based on a
fundamental tradeoff between the violation of Bell inequalities and the
unpredictability of quantum measurements. The simplest setting in which this
tradeoff can be stated involves two separate parties, Alice and Bob, sharing
two subsystems in an entangled state on which they perform, respectively, one
of two measurements $x, y \in \{0, 1\}$ yielding one of two outcomes
$a, b \in \{0, 1\}$. In this setting, the expectation value
\begin{equation}
  \label{eq:chsh_def}
  S = \sum_{abxy} (-1)^{a + b + xy} P(ab \mid xy)
\end{equation}
of the CHSH Bell correlator \cite{ref:ch1969}, where $P(ab \mid xy)$ denotes
the joint probabilities for outcomes $a,b$ given measurements $x,y$, implies
the fundamental lower bound
\begin{equation}
  \label{eq:hmin_chsh}
  H_{\rmin}(A \mid \rE)
  \geq 1 - \log_{2} \bigro{1 + \sqrt{2 - S^{2} / 4}}
\end{equation}
on the min entropy $H_{\rmin}(A \mid \rE)$ of Alice's outcome conditioned on
one of Alice's inputs (say, $x = 0$) and the quantum side information $\rE$
of any potential adversary. This relation is tight and is attained with
equality with the optimal attack described in \cite{ref:ab2007}.

Contrarily to other tradeoffs used in standard QKD, which assume some level
of trust and characterisation of the quantum systems, the bound
\eqref{eq:hmin_chsh} is device independent in the sense that it holds for any
quantum state $\rho_{\rABE}$ and measurement operators $\{M_{a \mid x}\}$ and
$\{M_{b \mid y}\}$ characterising Alice's and Bob's devices. The relation
\eqref{eq:hmin_chsh} was first derived in the context of DI-randomness
certification \cite{ref:pam2010} and has since featured as an ingredient in a
number of DI QKD security proofs \cite{ref:mpa2011,ref:pma2013,ref:vv2014}.

Since they are based on the violation of Bell inequalities, DI QKD protocols
are entanglement-based (EB) protocols. Indeed, in the DI setting,
entanglement is necessary to guarantee security with a minimal set of
assumptions on the devices \cite{ref:pa2009}. Implementations of traditional
(non-DI) QKD protocols, such as BB84 \cite{ref:bb1984}, are, however, usually
of the prepare-and-measure (PM) type. In a PM protocol, Alice uses a source
to prepare certain states which are then transmitted through a quantum
channel to Bob who performs measurements on them. PM schemes have the
practical advantage that they do not require the manipulation of
entanglement. For this same reason, however, they cannot be fully DI. Recent
works have nevertheless considered the possibility of PM QKD schemes that are
at least \emph{partially} DI \cite{ref:pb2011,ref:p2015}.

In traditional QKD, a famous argument establishes an equivalence between the
security of PM and EB protocols \cite{ref:bbm1992}. In the BB84 protocol, for
instance, Alice could prepare the four BB84 states by preparing a $\Phi^{+}$
Bell state
$(\ket{0}_{\rA} \ket{0}_{\rA'} + \ket{1}_{\rA} \ket{1}_{\rA'}) / \sqrt{2}$
(in some Hilbert space $\hilb_{\rA} \otimes \hilb_{\rA'}$) in her lab and
measuring either in the computational ($\{\ket{0}_{\rA}, \ket{1}_{\rA}\}$)
basis or in the Hadamard ($\{\ket{+}_{\rA}, \ket{-}_{\rA}\}$) basis in
$\hilb_{\rA}$ and transmitting the projected state in $\hilb_{\rA'}$ to
Bob. Since the security can only be reduced if the $\Phi^{+}$ state is
replaced by a state $\ket{\psi}_{\rABE}$ chosen by an adversary (Eve) and
shared between Alice, Bob, and Eve (the situation considered in EB security
proofs), a security proof of the EB version of the BB84 protocol
automatically implies the security of the PM version.

\begin{figure}[btp]
  \centering
  \begin{tikzpicture}
  \def\diaglen{1.0607}

  \node[source] (alice) at (0,0) {$\rS_{\rA}$};
  \node[detector] (bob) at (8,0) {$\rM_{\rB}$};

  \draw node[above=2.5mm of alice] (alice_input) {$x, a$};
  \draw node[above=2.5mm of bob] (bob_input) {$y$};
  \draw node[below=2.5mm of bob] (bob_output) {$b$};

  \draw[post] (alice) -- node[below=2mm] {$\rho_{x,a}$} (bob);

  \draw[->] (alice_input) -- (alice);
  \draw[->] (bob_input) -- (bob);
  \draw[->] (bob) -- (bob_output);

  \begin{pgfonlayer}{background}
    \draw[dashed] ([upshift,rightshift]alice.north east)
      -- ([downshift,rightshift]alice.south east);
    \draw[dashed] ([upshift,leftshift]bob.north west)
      -- ([downshift,leftshift]bob.south west);
  \end{pgfonlayer}
\end{tikzpicture}
  \caption{Semi-device-independent scenario with the prepare-and-measure CHSH
    estimation. Alice's source ($\rS_{\rA}$) can emit one of four different
    qubit states $\rho_{x,a} \in \{\rho, \rho', \sigma, \sigma'\}$ depending
    on a choice of input $(x, a) \in \{0, 1\}^{2}$. Bob's measurement device
    ($\rM_{\rB}$) performs one of two measurements depending on a choice of
    input $y \in \{0, 1\}$, yielding an outcome $b \in \{0, 1\}$.}
  \label{fig:scenario}
\end{figure}
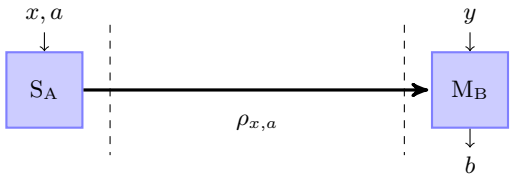

In the DI setting one can similarly associate a corresponding PM scheme to
any EB scheme. In particular, one can consider a PM version of the above CHSH
scenario, as illustrated in Fig.~\ref{fig:scenario}. In this PM version,
Alice possesses a source which can emit one of four different quantum states,
noted $\rho$, $\rho'$, $\sigma$, and $\sigma'$, depending on a respective
choice of input $(x, a) = (0, 0)$, $(0, 1)$, $(1, 0)$, and $(1, 1)$. Alice
randomly chooses $x \in \{0, 1\}$ (not necessarily equiprobably) and chooses
$a \in \{0, 1\}$ randomly and equiprobably and attempts to transmit the
corresponding state to Bob, who may perform one of two binary-outcome
measurements on them (indexed by the input $y \in \{0, 1\}$ and output
$b \in \{0, 1\}$). We can then define
\begin{equation}
  \label{eq:pmchsh_def}
  S = \frac{1}{2} \sum_{abxy} (-1)^{a + b + xy} P(b \mid axy) \,.
\end{equation}
as the PM analogue of the CHSH correlator \eqref{eq:chsh_def}.

In a traditional (non-DI) setting, the equivalence between the EB and PM
scenarios would imply that the bound \eqref{eq:hmin_chsh} on Alice's
randomness as a function of the CHSH correlator also holds in the PM
version. In a DI setting, however, this equivalence is not immediate at all.
First, the PM version cannot be fully DI (because the source could simply
transmit Alice's choice of input classically). The security of a PM version
will thus depend on some minimal assumption about the source. One possibility
is to assume a dimension bound on one or more of the devices; such
\emph{semi}-DI PM schemes were proposed in \cite{ref:pb2011}. Second, states
prepared by measurements on half of an entangled pair satisfy a constraint
called \emph{basis independence}: if a set $\{\rho_{x}\}$ of states is
prepared with associated probabilities $\{p_{x}\}$ by performing a
measurement on half an entangled pair, the average state $\sum_{x} p_{x}
\rho_{x}$ is independent of the measurement used to prepare it (a version of
the no-signalling principle). The basis-independence constraint, however,
need not be satisfied, and is actually explicitly relaxed, in the PM setting.

We show here that the fundamental bound on Alice's min entropy
\eqref{eq:hmin_chsh} nevertheless still holds in a semi-DI setting, with the
PM version of the CHSH correlator \eqref{eq:pmchsh_def} used in place of
\eqref{eq:chsh_def}. Such a result can then be used to bring the semi-DI
setting (for which security proofs are lacking) closer in line to known
security results for DI QKD. In particular, the conditional min entropy can,
for instance, be used to lower bound the Devetak-Winter key rate
\cite{ref:dw2005} in order to establish the security against collective
attacks of a semi-DI QKD protocol based on the estimation of the CHSH
correlator \eqref{eq:pmchsh_def}.

\paragraph*{Dimension assumption.---}
Let us start by making precise the assumption that we need to derive
\eqref{eq:hmin_chsh} in the PM setting. During the transmission from Alice to
Bob, an adversary may perform an arbitrary unitary operation on the states
sent by Alice, with the intent of gaining some information about them.  (More
general quantum operations can be made unitary by enlarging the adversary's
Hilbert space, according to Stinespring's dilation theorem.)  Following this
unitary attack, the emitted state $\rho_{x,a}$ is now shared between Bob and
Eve \footnote{In an abuse of notation, we will write $\rho_{x,a}$ to refer
  both to the states leaving Alice's device and to the states after Eve's
  unitary attack.}, i.e., acts on a Hilbert space $\hilb_{\rB} \otimes
\hilb_{\rE}$. We make the assumption that the two differences $\rho - \rho'$
and $\sigma - \sigma'$ between the source states (after the unitary attack)
share their support on a common two-dimensional subspace $\hilb_{\rA}$ of
$\hilb_{\rB} \otimes \hilb_{\rE}$. We refer to this condition as the
\emph{qubit source assumption}. We will later discuss the physical
implications of this assumption; for now we simply take it as a mathematical
condition satisfied by the states prepared by Alice's box.

A simple example illustrates the necessity of the qubit source
assumption. Specifically, if Alice's source prepares pure states
$\rho_{x,a} = \proj{\psi_{x,a}}$ which are duplicate copies of the BB84
states,
\begin{IEEEeqnarray}{rCl+rCl}
  \ket{\psi_{00}} &=& \ket{0}_{\rB} \ket{0}_{\rE} \,, &
  \ket{\psi_{01}} &=& \ket{1}_{\rB} \ket{1}_{\rE} \,, \\
  \ket{\psi_{10}} &=& \ket{+}_{\rB} \ket{+}_{\rE} \,, &
  \ket{\psi_{11}} &=& \ket{-}_{\rB} \ket{-}_{\rE} \,,
\end{IEEEeqnarray}
in which $\ket{0}$ and $\ket{1}$ are orthonormal and
$\ket{\pm} = \tfrac{1}{\sqrt{2}}[\ket{0} \pm \ket{1}]$, the maximal value
$S = 2 \sqrt{2}$ can be attained while Eve always acquires exactly the same
state as Bob. These states are not linearly independent (one can readily
verify that
$\ket{0} \ket{0} + \ket{1} \ket{1} = \ket{+} \ket{+} + \ket{-} \ket{-}$) and
span a three-dimensional Hilbert space, from which we see that the security
of the semi-DI scenario is fully compromised if the qubit source assumption
is not satisfied.

\paragraph*{Min entropy and Eve's distinguishability.---}
To prove the bound \eqref{eq:hmin_chsh}, let us first note that if the input
$a$ is chosen equiprobably, its min entropy, conditioned on the case $x = 0$
and on Eve's quantum side information, is a function of the classical-quantum
state
\begin{equation}
  \label{eq:cq_state}
  \tau_{A\rE} = \tfrac{1}{2} \proj{0} \otimes \rho\0_{\rE}
  + \tfrac{1}{2} \proj{1} \otimes \rho\1_{\rE} \,,
\end{equation}
in which $\rho\0_{\rE}$ and $\rho\1_{\rE}$ are Eve's marginals of the states
$\rho$ and $\rho'$ after some given unitary attack (in the rest of this
article, subscripts indicate partial tracing in the obvious way, e.g.,
$\rho_{\rB} = \Tr_{\rE}[\rho]$). Evaluated on \eqref{eq:cq_state}, the
min entropy can be expressed \cite{ref:krs2009,ref:wlp2013} as
\begin{equation}
  \label{eq:hmin_trdist}
  H_{\rmin}(A \mid \rE) = 1 - \log_{2} \bigro{
    1 + D(\rho\0_{\rE}, \rho\1_{\rE})} \,,
\end{equation}
with the trace distance between $\rho\0_{\rE}$ and $\rho\1_{\rE}$ defined by
$D(\rho\0_{\rE}, \rho\1_{\rE}) = \tfrac{1}{2} \trnorm{\rho\0_{\rE} -
  \rho\1_{\rE}}$
where $\trnorm{A} = \Tr[\sqrt{A^{\dagger} A}]$ denotes the trace norm of an
operator $A$. We will obtain the main result \eqref{eq:hmin_chsh} by showing
that the trace distance appearing in \eqref{eq:hmin_trdist} is upper bounded
by
\begin{equation}
  \label{eq:trdist_S_bound}
  D(\rho\0_{\rE}, \rho\1_{\rE}) \leq \sqrt{2 - S^{2} / 4}
\end{equation}
in terms of $S$.

\paragraph*{Outline of proof.---}
We now outline the derivation of \eqref{eq:trdist_S_bound}. The lengthier and
more pedestrian parts of the proof are given in the appendix.

Let us first introduce operators $Z$ and $X$ defined such
that
\begin{IEEEeqnarray}{rCl}
  \label{eq:Z_def}
  \rho - \rho' &=& \alpha Z \,, \\
  \label{eq:X_def}
  \sigma - \sigma' &=& \beta X \,, 
\end{IEEEeqnarray}
with $\alpha = \tfrac{1}{2} \trnorm{\rho - \rho'}$ and
$\beta = \tfrac{1}{2} \trnorm{\sigma - \sigma'}$ such that the (traceless
qubit) operators $Z$ and $X$ satisfy
$\tfrac{1}{2} \trnorm{Z} = \tfrac{1}{2} \trnorm{X} = 1$. Then, in terms of
these operators, the CHSH expectation value \eqref{eq:pmchsh_def} can be
expressed as
\begin{equation}
  \label{eq:chsh_cc_trexpr}
  S = \tfrac{1}{2} \Tr \bigsq{
    U_{\rB} \bigro{\alpha Z_{\rB} + \beta X_{\rB}} 
    + V_{\rB} \bigro{\alpha Z_{\rB} - \beta X_{\rB}}} \,,
\end{equation}
where $U_{\rB} = \sum_{b}(-1)^{b} M_{b \mid y=0}$ and
$V_{\rB} = \sum_{b}(-1)^{b} M_{b \mid y=1}$ are Hermitian unitary operators
acting on $\hilb_{\rB}$ describing the observables corresponding to the
(without loss of generality, projective) measurements $y = 0$ and $y = 1$.

A general result for any pair of Hermitian unitaries is that they admit a
common block diagonalisation in blocks of dimension no more than 2. We can
thus set
\begin{IEEEeqnarray}{rCl+rCl}
  \label{eq:UV_block_diag}
  U_{\rB} &=& \bigoplus_{k} U^{k}_{\rB} \,, &
  V_{\rB} &=& \bigoplus_{k} V^{k}_{\rB} \,,
\end{IEEEeqnarray}
in which $U^{k}_{\rB}$ and $V^{k}_{\rB}$ are still Hermitian and unitary and
of dimension at most 2, $\forall k$ (the Jordan lemma \cite{ref:j1875}, see
Lemma~2 of \cite{ref:pa2009} for a short proof). This reduces the problem to
considering qubit subspaces on Bob's side. For each subspace $k$, we can
define the corresponding contribution to $S$ by
\begin{equation}
  \label{eq:S_k_def}
  S_{k} = \tfrac{1}{2} \alpha \Tr[(U^{k}_{\rB} + V^{k}_{\rB}) Z_{\rB}]
  + \tfrac{1}{2} \beta \Tr[(U^{k}_{\rB} - V^{k}_{\rB}) X_{\rB}] \,,
\end{equation}
with $\sum_{k} S_{k} = S$. Similarly, we define a probabilistic weight for
each subspace $k$ by
\begin{equation}
  \label{eq:p_k_def}
  p_{k} = \tfrac{1}{2} \Tr[\id^{k}_{\rB} \mathcal{I}_{\rB}] \,,
\end{equation}
with $\sum_{k} p_{k} = 1$, defined in terms of the projection operator
$\id^{k}_{\rB}$ on the $k$th subspace (satisfying
$\id^{k}_{\rB} = (U^{k}_{\rB})^{2} = (V^{k}_{\rB})^{2}$) and the partial
trace $\mathcal{I}_{\rB} = \Tr_{\rE}[\mathcal{I}]$ of the identity on the
space of source states (satisfying $\mathcal{I} = Z^{2} = X^{2}$).

We now introduce an orthonormal basis $\{\ket{\ry}, \ket{\ry'}\}$ of the
space of source states chosen such that $Y = \proj{\ry} - \proj{\ry'}$ is
orthogonal to the operators $Z$ and $X$, defined by Eqs.~\eqref{eq:Z_def} and
\eqref{eq:X_def}, on the Bloch sphere. In this basis, in an appropriate phase
convention, $Z$ and $X$ can be expressed as
\begin{IEEEeqnarray}{rCl}
  \label{eq:Z_y}
  Z &=& \ee^{\ii \tfrac{\varphi}{2}} \trans{\ry}{\ry'}
  + \ee^{-\ii \tfrac{\varphi}{2}} \trans{\ry'}{\ry} \,, \\
  \label{eq:X_y}
  X &=& \ee^{-\ii \tfrac{\varphi}{2}} \trans{\ry}{\ry'}
  + \ee^{\ii \tfrac{\varphi}{2}} \trans{\ry'}{\ry}
\end{IEEEeqnarray}
for some (\emph{a priori} unknown) angle $\varphi$, while the source space
identity operator $\mathcal{I}$ takes the expression
\begin{equation}
  \mathcal{I} = \proj{\ry} + \proj{\ry'} \,.
\end{equation}
One can readily verify that $\acomm{Z}{Y} = \acomm{X}{Y} = 0$, that
$\comm{Z}{X} = 2 i \sin(\varphi) Y$, and that $\acomm{Z}{X} = \cos(\varphi)
\mathcal{I}$. An important step in the derivation of the trace-distance bound
\eqref{eq:trdist_S_bound} consists in turning the value of $S$ into a
constraint on the part $Y_{\rB}$ of the operator $Y$ accessible to Bob
\footnote{Roughly, if we think of Alice's states as living on the $\sz-\sx$
  plane, we want to show that an attack that doesn't degrade $S$ too much
  also can't affect the $\sy$ basis too much.}. Specifically, in each
subspace $k$ defined by the block diagonalisation \eqref{eq:UV_block_diag},
we prove in Appendix~\ref{sec:tr_WkY_S_proof} that there exists a Hermitian
unitary operator $W^{k}_{\rB}$ with the property that
\begin{equation}
  \label{eq:tr_WkY_S_bound}
  \alpha \tfrac{1}{2} \Tr[W^{k}_{\rB} Y_{\rB}]
  \geq \sqrt{S\subsup{k}{2} / 4 - p\subsup{k}{2}} \,,
\end{equation}
where $S_{k}$ and $p_{k}$ are as defined in \eqref{eq:S_k_def} and
\eqref{eq:p_k_def}. Eq.~\eqref{eq:tr_WkY_S_bound} holds regardless of the
value of $\beta$ appearing in \eqref{eq:S_k_def} and of $\varphi$ in
\eqref{eq:Z_y} and \eqref{eq:X_y}. Note that $\alpha$ can also be eliminated
using that $\tfrac{1}{2} \Tr[W^{k}_{\rB} Y_{\rB}] \geq \alpha \tfrac{1}{2}
\Tr[W^{k}_{\rB} Y_{\rB}]$.

In order to obtain the upper bound \eqref{eq:trdist_S_bound} on
$D(\rho\0_{\rE}, \rho\1_{\rE})$, we also derive a tradeoff between the
quantity $\tfrac{1}{2} \Tr[W^{k}_{\rB} Y_{\rB}]$, appearing in
\eqref{eq:tr_WkY_S_bound}, and the distinguishability of Eve's states.
Specifically, we prove in Appendix~\ref{sec:Z2_Y2_k_proof} that, for any
Hermitian unitary $U_{\rE}$ acting on $\hilb_{\rE}$, the inequality
\begin{equation}
  \label{eq:Z2_Y2_k}
  \tfrac{1}{4} \Tr[W^{k}_{\rB} Y_{\rB}]^{2}
  + \tfrac{1}{4} \Tr \bigsq{(\id^{k}_{\rB} \otimes U_{\rE}) Z}^{2}
  \leq p\subsup{k}{2}
\end{equation}
holds in each subspace $k$.

We obtain \eqref{eq:trdist_S_bound} by taking for $U_{\rE}$ in
\eqref{eq:Z2_Y2_k} a Hermitian unitary such that
$\tfrac{1}{2} \Tr[U_{\rE} Z_{\rE}] = \tfrac{1}{2} \trnorm{Z_{\rE}}$
\footnote{Such a unitary can be constructed explicitly by diagonalising
  $Z_{\rE}$. Specifically, if
  \unexpanded{$Z_{\rE} = \sum_{k} \lambda_{k} \proj{k}$} is an expression for
  $Z_{\rE}$ in diagonal form, then, for instance, $U_{\rE}$ can be taken to
  be \unexpanded{$U_{\rE} = \sum_{k \in K_{\geq}} \proj{k} - \sum_{k \in
      K_{<}} \proj{k}$} where $K_{\geq} = \{k : \lambda_{k} \geq 0\}$ and
  $K_{<} = \{k : \lambda_{k} < 0\}$ are the sets of indices corresponding,
  respectively, to the nonnegative and negative eigenvalues of
  $Z_{\rE}$.}. Because $D(\rho\0_{\rE}, \rho\1_{\rE}) = \alpha \tfrac{1}{2}
\trnorm{Z_{\rE}} \leq \tfrac{1}{2} \trnorm{Z_{\rE}}$, the trace distance is
upper bounded by
\begin{equation}
  \label{eq:trdist_Zk}
  D(\rho\0_{\rE}, \rho\1_{\rE})
  \leq \sum_{k} \tfrac{1}{2} \Tr \bigsq{
    (\id^{k}_{\rB} \otimes U_{\rE}) Z} \,.
\end{equation}
Using \eqref{eq:tr_WkY_S_bound} and \eqref{eq:Z2_Y2_k} and omitting $\alpha$,
we have
\begin{equation}
  \label{eq:trdist_S_k}
  \tfrac{1}{2} \Tr \bigsq{(\id^{k}_{\rB} \otimes U_{\rE}) Z}
  \leq p_{k} \sqrt{2 - (S_{k} / p_{k})^{2} / 4} \,,
\end{equation}
and substituting \eqref{eq:trdist_S_k} into \eqref{eq:trdist_Zk} and using
that the function $S \mapsto \sqrt{2 - S^{2} / 4}$ is concave, we finally
obtain
\begin{IEEEeqnarray}{rCl}
  D(\rho\0_{\rE}, \rho\1_{\rE})
  &\leq& \sum_{k} \tfrac{1}{2} \Tr \bigsq{(\id^{k}_{\rB} \otimes U_{\rE}) Z}
  \IEEEnonumber \\
  &\leq& \sum_{k} p_{k} \sqrt{2 - (S_{k}/p_{k})^{2} / 4} \IEEEnonumber \\
  &\leq& \sqrt{2 - \bigro{\textstyle{\sum_{k} S_{k}}}^{2} / 4}
  \IEEEnonumber \\
  &=& \sqrt{2 - S^{2} / 4} \,.
\end{IEEEeqnarray}
Combining with the expression \eqref{eq:hmin_trdist} for the min entropy, we
obtain \eqref{eq:hmin_chsh}.

As with its EB counterpart, \eqref{eq:trdist_S_bound} and the resulting
min-entropy bound are tight and are attained with a PM version of the optimal
attack originally given in \cite{ref:ab2007}; for completeness we have
included a description of this attack in Appendix~\ref{sec:bound_tightness}.

\paragraph*{Discussion of the qubit assumption.---}
Having proven our main result, let us now discuss the qubit source assumption
in more detail. Note first that Alice's ``preparation'' device may not in
general actually prepare a new state from scratch, but instead implement a
transformation on a preexisting qubit stored in her box, which could be
entangled with Eve's system prior to the protocol. The presence of such prior
entanglement between Alice's device and Eve may completely break the security
of a PM scheme, as noted in \cite{ref:pb2011}. However, since our qubit
assumption is formulated in the total space $\hilb_{\rB} \otimes \hilb_{\rE}$
including Eve's Hilbert space, it naturally limits the amount of potential
prior entanglement between Alice and Eve (or Alice and Bob) and thus a nice
mathematical feature of our formulation is that we do not need to state this
limitation on prior entanglement as a separate, additional assumption.

On the other hand, since our qubit assumption is formulated in the space
$\hilb_{\rB} \otimes \hilb_{\rE}$ after Eve's attack, it may not be possible
to practically verify this assumption in a cryptographic setting (since Alice
and Bob do not have access to Eve's system). Note, however, that a sufficient
condition for our assumption to be satisfied is that
\begin{inparaenum}[(i)]
  \item there exists no prior entanglement between Alice and Eve or Bob
  (e.g., Alice's preparation box has no quantum memory), and
  \item the states sent by Alice's box, before going through the channel and
  suffering Eve's (without loss of generality, unitary) attack, are such that
  $\rho - \rho'$ and $\sigma - \sigma'$ have support in the same
  two-dimensional subspace.
\end{inparaenum}
Under these conditions, the states $\rho - \rho'$ and $\sigma - \sigma'$
after Eve's unitary attack will still share the same two-dimensional support
and thus our qubit source assumption will be satisfied. However, the
condition that we use to derive the min-entropy bound \eqref{eq:hmin_chsh} is
formally weaker than the combination of $i)$ and $ii)$ as these only
represent sufficient conditions for our assumption to be satisfied.

Another nice feature of our formulation is that the qubit assumption refers
only to the differences $\rho - \rho'$ and $\sigma - \sigma'$ and not
directly to the states $\rho_{x, a}$ themselves, which may live in a higher
dimensional Hilbert space. For instance, in an optical implementation, each
``qubit'' may be a qubit encoded in the polarisation degree of freedom of a
single photon, but may also possess a vacuum component and thus formally be a
three-level system of the form $\rho_{x, a} = p \proj{0} + (1 - p)
\tilde{\rho}_{a, x}$, where $\tilde{\rho}_{a, x}$ is the one-photon polarised
qubit part. Still, the differences $\rho - \rho'$ and $\sigma - \sigma'$ only
involve the genuine qubit parts and thus satisfy our qubit source
assumption.

Finally, let us remark that our assumption can immediately be weakened in two
ways. First, using convexity arguments, it is easy to see that the
min-entropy bound \eqref{eq:hmin_chsh} still holds if Alice's, Bob's, and
Eve's systems share prior classical randomness, provided that for any value
$\lambda$ of the shared randomness, the differences
$\rho\0_{\lambda} - \rho\1_{\lambda}$ and
$\sigma\0_{\lambda} - \sigma\1_{\lambda}$ satisfy the qubit
assumption. Again, it may not be possible to practically verify this
assumption in the most general DI setting (as Alice and Bob will not have
access to the individual values of the shared randomness if their devices are
uncharacterised). However, a sufficient condition for this assumption to be
satisfied is if each of the averaged states
$\rho_{x, a} = \sum_{\lambda} q_{\lambda} \rho_{x, a; \lambda}$ are contained
in the same qubit space, a condition which does not require any knowledge of
the shared randomness.

Second, we point out that the bound on the min entropy is also robust with
respect to the qubit assumption; i.e., this assumption need only be
approximately verified. Specifically, suppose that, instead of assuming
\eqref{eq:Z_def} and \eqref{eq:X_def}, we assume that there exist traceless
two-dimensional unit operators $\alpha \tilde{Z}$ and $\beta \tilde{X}$ such
that $\frac{1}{2}\trnorm{(\rho - \rho') - \alpha \tilde{Z}} \leq \varepsilon$
and $\frac{1}{2} \trnorm{(\sigma - \sigma') - \beta \tilde{X}} \leq
\varepsilon$. Then it is easy to see that $D(\rho\0_{\rE}, \rho\1_{\rE}) \leq
\tfrac{1}{2} \trnorm{\alpha \tilde{Z}_{\rE}} + \varepsilon$ and that the CHSH
expectation value computed with $\alpha \tilde{Z}$ and $\beta \tilde{X}$
cannot differ from $S$ by more than $4 \varepsilon$. Small deviations from
the qubit source assumption can thus be tolerated, with a bound on the
min entropy no worse than
\begin{equation}
  H_{\rmin}(A \mid \rE) \geq 1 - \log_{2} \bigro{
    1 + \sqrt{2 - (S - 4 \varepsilon)^{2} / 4} + \varepsilon} \,.
\end{equation}

\paragraph*{Conclusion.---}
We gave a proof that the fundamental lower bound \eqref{eq:hmin_chsh} on the
randomness of Alice's outcomes as a function of the CHSH expression,
originally derived in the context of device-independent QKD and randomness
certification, still holds in a PM setting with a qubit assumption. Though
the equivalence between EB and PM schemes in standard QKD may \emph{a priori}
suggest that this should naturally be the case, this is not at all immediate
as this equivalence breaks in a DI setting. Indeed, the techniques that we
have used here to establish the lower bound \eqref{eq:hmin_chsh} in the PM
setting are quite different from the ones used to establish the EB version of
this bound.

This fundamental lower bound \eqref{eq:hmin_chsh} can now, in principle, be
used as a building block to prove the security of semi-DI QKD protocols, in
the same way that it was used in the fully DI setting in
Refs.~\cite{ref:mpa2011,ref:pma2013,ref:vv2014}.

We remark that in the EB scenario, an analogous tight bound for the Holevo
quantity (or, equivalently, the conditional von Neumann entropy) instead of
the min entropy had earlier been presented in \cite{ref:ab2007} as part of a
security proof against collective attacks. The conditional von Neumann
entropy can likewise, in principle, be bounded in the PM scenario. A partial
result for the von Neumann entropy, restricted to the case where Bob's
measurements are additionally assumed to be two-dimensional, is given in
Ref.~\cite{ref:w2014b}.

Finally, having shown that our min-entropy bound holds for Alice's system
conditioned on Eve, it would be interesting to investigate whether a similar
result holds for the min entropy $H_{\rmin}(B \mid \rE)$ associated with
Bob's measurement outcome. In particular, a version of this result
conditioned on just one of Alice's state preparations would apply immediately
to the problem of randomness certification
\cite{ref:c2006,ref:pam2010,ref:vv2012,ref:ms2014}, which has similarly been
investigated in PM scenarios \cite{ref:ly2011,ref:lb2015}.

\begin{acknowledgments}
  This work is supported by the Spanish project FOQUS, the Generalitat de
  Catalunya (SGR 875), the EU projects QITBOX and QALGO, the F.R.S.-FNRS
  under the project DIQIP, and by the Interuniversity Attraction Poles
  program of the Belgian Science Policy Office under the grant IAP P7-35
  photonics@be. S.~P. is a Research Associate of the Fonds de la Recherche
  Scientifique F.R.S.-FNRS (Belgium). E.~W. was supported by a Belgian Fonds
  pour la Formation \`a{} la Recherche dans l'Industrie et dans l'Agriculture
  (F.R.I.A.) grant while this work was started.
\end{acknowledgments}

\appendix

\section{Proof of \eqref{eq:tr_WkY_S_bound}}
\label{sec:tr_WkY_S_proof}

As explained in the main text, after applying the Jordan lemma the optimal
expectation value \eqref{eq:chsh_cc_trexpr} of $S$ can be expressed as
$S = \sum_{k} S_{k}$, with
\begin{equation}
  S_{k} = \tfrac{1}{2} \alpha \Tr[(U_{k} + V_{k}) Z]
  + \tfrac{1}{2} \beta \Tr[(U_{k} - V_{k}) X] \,,
\end{equation}
where we have set $U_{k} = U^{k}_{\rB} \otimes \id_{\rE}$ and $V_{k} =
V^{k}_{\rB} \otimes \id_{\rE}$, and $U^{k}_{\rB}$ and $V^{k}_{\rB}$ are the
Hermitian unitary operators of dimension at most 2 appearing in
\eqref{eq:UV_block_diag} and \eqref{eq:S_k_def}.

Inserting the expressions \eqref{eq:Z_y} and \eqref{eq:X_y} for the operators
$Z$ and $X$, $S_{k}$ can be expressed as
\begin{IEEEeqnarray}{rCl}
  \label{eq:S_k_yUV}
  S_{k} &=& \alpha \re \bigsq{\bra{\ry} \ee^{-\ii \tfrac{\varphi}{2}}
    (U_{k} + V_{k}) \ket{\ry'}} \IEEEnonumber \\
  &&+\> \beta \re \bigsq{\bra{\ry} \ee^{\ii \tfrac{\varphi}{2}}
    (U_{k} - V_{k}) \ket{\ry'}}
\end{IEEEeqnarray}
in terms of the $\ry$ states. The only interesting case is where both
$U^{k}_{\rB}$ and $V^{k}_{\rB}$ are two-dimensional and of eigenvalues $+1$
and $-1$. In this case, if $U^{k}_{\rB}$ and $V^{k}_{\rB}$ are separated by
an angle $\gamma_{k}$ on the Bloch sphere, one can choose an orthonormal
basis $\{\ket{\rw_{k}}_{\rB}, \ket{\rw_{k}'}_{\rB}\}$ in which
\begin{IEEEeqnarray}{rCl}
  U^{k}_{\rB} + V^{k}_{\rB} &=& 2 \cos \bigro{\tfrac{\gamma_{k}}{2}}
  \bigro{\trans{\rw_{k}}{\rw'_{k}}_{\rB}
    + \trans{\rw'_{k}}{\rw_{k}}_{\rB}} \,, \\
  U^{k}_{\rB} - V^{k}_{\rB} &=& 2 \sin \bigro{\tfrac{\gamma_{k}}{2}}
  \bigro{i \trans{\rw_{k}}{\rw'_{k}}_{\rB}
    - i \trans{\rw'_{k}}{\rw_{k}}_{\rB}} \,. \IEEEeqnarraynumspace
\end{IEEEeqnarray}
Inserting this into \eqref{eq:S_k_yUV}, the expression for $S_{k}$ can be
simplified to
\begin{equation}
  S_{k} / 2 = \re \bigsq{\bra{\ry} \bigro{
      \lambda_{k} \trans{\rw_{k}}{\rw'_{k}}_{\rB}
      + \mu_{k} \trans{\rw'_{k}}{\rw_{k}}_{\rB}}
    \otimes \id_{\rE} \ket{\ry'}} \,,
\end{equation}
where we have collected the various angles into
\begin{IEEEeqnarray}{rCl}
  \lambda_{k} &=& \alpha
  \cos \bigro{\tfrac{\gamma_{k}}{2}} \ee^{-\ii \tfrac{\varphi}{2}}
  + \ii \beta \sin \bigro{\tfrac{\gamma_{k}}{2}}
  \ee^{\ii \tfrac{\varphi}{2}} \,, \IEEEnonumber \\
  \mu_{k} &=& \alpha
  \cos \bigro{\tfrac{\gamma_{k}}{2}} \ee^{-\ii \tfrac{\varphi}{2}}
  -\ii \beta \sin \bigro{\tfrac{\gamma_{k}}{2}}
  \ee^{\ii \tfrac{\varphi}{2}}
\end{IEEEeqnarray}
for convenience. Introducing now vectors
\begin{IEEEeqnarray}{rCl}
  \ket{\rA_{k}}
  &=& (\bra{\rw_{k}}_{\rB} \otimes \id_{\rE}) \ket{\ry} \,, \\
  \ket{\rA'_{k}}
  &=& (\bra{\rw'_{k}}_{\rB} \otimes \id_{\rE}) \ket{\ry'} \,, \\
  \ket{\rB_{k}}
  &=& (\bra{\rw'_{k}}_{\rB} \otimes \id_{\rE}) \ket{\ry} \,, \\
  \ket{\rB'_{k}}
  &=& (\bra{\rw_{k}}_{\rB} \otimes \id_{\rE}) \ket{\ry'}
\end{IEEEeqnarray}
(in $\hilb_{\rE}$) in order to further simplify the notation, we obtain
\begin{equation}
  S_{k} / 2 = \re \bigsq{ \lambda_{k} \braket{\rA_{k}}{\rA'_{k}}
    + \mu_{k} \braket{\rB_{k}}{\rB'_{k}}} \,.
\end{equation}
Finally, in order to reexpress $S_{k}$ in a form better suited for
determining an upper bound, we introduce new coefficients
\begin{IEEEeqnarray}{rCl}
  \xi_{k} &=& \cos \bigro{\tfrac{\gamma_{k} + \varphi}{2}}
  + \ii \sin \bigro{\tfrac{\gamma_{k} - \varphi}{2}} \,, \\
  \nu_{k} &=& \cos \bigro{\tfrac{\gamma_{k} - \varphi}{2}}
  - \ii \sin \bigro{\tfrac{\gamma_{k} + \varphi}{2}} \,,
\end{IEEEeqnarray}
and
\begin{equation}
  \gamma_{\pm} = \tfrac{1}{2} (\alpha \pm \beta) \,,
\end{equation}
such that
\begin{IEEEeqnarray}{rCl}
  \lambda_{k} &=& \gamma_{+} \xi_{k} + \gamma_{-} \nu_{k} \,, \\
  \mu_{k} &=& \gamma_{+} \nu_{k} + \gamma_{-} \xi_{k} \,.
\end{IEEEeqnarray}
Inserting these into the expression for $S_{k}$, we arrive at
\begin{IEEEeqnarray}{rl}
  \label{eq:S_k_eq_AA_BB}
  S_{k} / 2 = \re \bigsq{& \xi_{k} \bigro{
      \gamma_{+} \braket{\rA_{k}}{\rA'_{k}}
      + \gamma_{-} \braket{\rB_{k}}{\rB'_{k}}} \IEEEnonumber \\
    &+\> \nu_{k} \bigro{\gamma_{-} \braket{\rA_{k}}{\rA'_{k}}
      + \gamma_{+} \braket{\rB_{k}}{\rB'_{k}}}} \,. \IEEEeqnarraynumspace
\end{IEEEeqnarray}

In order to obtain a useful upper bound on $S_{k}$, we begin by taking the
absolute value of the various terms in \eqref{eq:S_k_eq_AA_BB}, obtaining
\begin{IEEEeqnarray}{rCl}
  S_{k} / 2 &\leq& \abs{\xi_{k}} \bigro{
    \abs{\gamma_{+}} \abs{\braket{\rA_{k}}{\rA'_{k}}}
    + \abs{\gamma_{-}} \abs{\braket{\rB_{k}}{\rB'_{k}}}} \IEEEnonumber \\
  &&+\> \abs{\nu_{k}} \bigro{
    \abs{\gamma_{-}} \babs{\braket{\rA_{k}}{\rA'_{k}}}
    + \abs{\gamma_{+}} \babs{\braket{\rB_{k}}{\rB'_{k}}}}
  \,. \IEEEeqnarraynumspace
\end{IEEEeqnarray}
Applying the Cauchy-Schwarz inequality, using that $\abs{\xi_{k}}^{2} +
\abs{\nu_{k}}^{2} = 2$, and developing,
\begin{IEEEeqnarray}{rCl}
  S\subsup{k}{2} / 4 &\leq& 2 \Bigro{
    \abs{\gamma_{+}} \babs{\braket{\rA_{k}}{\rA'_{k}}}
    + \abs{\gamma_{-}} \babs{\braket{\rB_{k}}{\rB'_{k}}}}^{2}
  \IEEEnonumber \\
  &&+\> 2 \Bigro{
    \abs{\gamma_{-}} \babs{\braket{\rA_{k}}{\rA'_{k}}}
    + \abs{\gamma_{+}} \babs{\braket{\rB_{k}}{\rB'_{k}}}}^{2}
  \IEEEnonumber \\
  &=& 2 (\gamma\subsup{+}{2} + \gamma\subsup{-}{2}) \Bigro{
    \babs{\braket{\rA_{k}}{\rA'_{k}}}^{2}
    + \babs{\braket{\rB_{k}}{\rB'_{k}}}^{2}} \IEEEnonumber \\
  &&+\> 8 \abs{\gamma_{+} \gamma_{-}} \babs{\braket{\rA_{k}}{\rA'_{k}}}
  \babs{\braket{\rB_{k}}{\rB'_{k}}} \IEEEnonumber \\
  &\leq& 2 (\gamma\subsup{+}{2} + \gamma\subsup{-}{2}) \Bigro{
    \norm{\rA_{k}}^{2} \norm{\rA'_{k}}^{2}
    + \norm{\rB_{k}}^{2} \norm{\rB'_{k}}^{2}} \IEEEnonumber \\
  &&+\> 8 \abs{\gamma_{+} \gamma_{-}} \norm{\rA_{k}} \norm{\rA'_{k}}
  \norm{\rB_{k}} \norm{\rB'_{k}} \,,
\end{IEEEeqnarray}
where we used the Cauchy-Schwarz inequality again to substitute
$\abs{\braket{\rA_{k}}{\rA'_{k}}} \leq \norm{\rA_{k}} \norm{\rA'_{k}}$ and
$\abs{\braket{\rB_{k}}{\rB'_{k}}} \leq \norm{\rB_{k}}
\norm{\rB'_{k}}$.
Applying now that
\begin{IEEEeqnarray}{rCcCl}
  2 \norm{\rA_{k}} \norm{\rA'_{k}} &\leq& \Sigma_{\rA}
  &=& \norm{\rA_{k}}^{2} + \norm{\rA'_{k}}^{2} \,, \\
  2 \norm{\rB_{k}} \norm{\rB'_{k}} &\leq& \Sigma_{\rB}
  &=& \norm{\rB_{k}}^{2} + \norm{\rB'_{k}}^{2} \,,
\end{IEEEeqnarray}
we find that
\begin{IEEEeqnarray}{rCl}
  \label{eq:S_k_A2_B2_bound}
  S\subsup{k}{2} / 4 &\leq& \tfrac{1}{2} (\gamma\subsup{+}{2}
  + \gamma\subsup{-}{2}) \bigsq{\Sigma\subsup{\rA}{2}
    + \Sigma\subsup{\rB}{2}}
  + 2 \abs{\gamma_{+} \gamma_{-}} \Sigma_{\rA} \Sigma_{\rB} \IEEEnonumber \\
  &=& \min \bigro{\alpha^{2}, \beta^{2}} \tfrac{1}{4}
  \bigro{\Sigma_{\rA} - \Sigma_{\rB}}^{2} \IEEEnonumber \\
  &&+\> \max \bigro{\alpha^{2}, \beta^{2}}
  \tfrac{1}{4} \bigro{\Sigma_{\rA} + \Sigma_{\rB}}^{2} \,.
\end{IEEEeqnarray}
Reinserting the definitions of the vectors $\ket{\rA_{k}}$, $\ket{\rA'_{k}}$,
$\ket{\rB_{k}}$, and $\ket{\rB'_{k}}$, note that
\begin{IEEEeqnarray}{rCl}
  \label{eq:A2_idWY}
  \Sigma_{\rA} - \Sigma_{\rB} &=& \Tr[W^{k}_{\rB} Y_{\rB}] \,, \\
  \label{eq:B2_idWY}
  \Sigma_{\rA} + \Sigma_{\rB} &=& \Tr[\id^{k}_{\rB} \mathcal{I}_{\rB}] \,,
\end{IEEEeqnarray}
where we recall that we defined
\begin{IEEEeqnarray}{rCl}
  \mathcal{I} &=& \proj{\ry} + \proj{\ry'} \,, \\
  Y &=& \proj{\ry} - \proj{\ry'} \,,
\end{IEEEeqnarray}
$\mathcal{I}_{\rB} = \Tr_{\rE}[\mathcal{I}]$, $Y_{\rB} = \Tr_{\rE}[Y]$, and
we have introduced
\begin{IEEEeqnarray}{rCl}
  \label{eq:idkB_def}
  \id^{k}_{\rB} &=& \proj{\rw_{k}}_{\rB} + \proj{\rw'_{k}}_{\rB} \,, \\
  \label{eq:WkB_def}
  W^{k}_{\rB} &=& \proj{\rw_{k}}_{\rB} - \proj{\rw'_{k}}_{\rB} \,.
\end{IEEEeqnarray}
In this way, we find that $S_{k}$ is upper bounded by
\begin{IEEEeqnarray}{rCl}
  \label{eq:S_k_tight_bound}
  S\subsup{k}{2} / 4 &\leq& \min \bigro{\alpha^{2}, \beta^{2}}
  \tfrac{1}{4} \Tr[W^{k}_{\rB} Y_{\rB}]^{2} \IEEEnonumber \\
  &&+\> \max \bigro{\alpha^{2}, \beta^{2}}
  \tfrac{1}{4} \Tr[\id^{k}_{\rB} \mathcal{I}_{\rB}]^{2} \,.
\end{IEEEeqnarray}
Finally, we substitute $\min(\alpha^{2}, \beta^{2}) \leq \alpha^{2}$,
$\max(\alpha^{2}, \beta^{2}) \leq 1$, and $p_{k} = \tfrac{1}{2}
\Tr[\id^{k}_{\rB} \mathcal{I}_{\rB}]$ in order to obtain
\begin{equation}
  \label{eq:S_k_bound}
  S\subsup{k}{2} / 4 \leq
  \alpha^{2} \tfrac{1}{4} \Tr[W^{k}_{\rB} Y_{\rB}]^{2} + p\subsup{k}{2} \,,
\end{equation}
which rearranges to \eqref{eq:tr_WkY_S_bound} in the main text. (If
$\tfrac{1}{2} \Tr[W^{k}_{\rB} Y_{\rB}]$ is negative, we simply replace
$W^{k}_{\rB} \mapsto - W^{k}_{\rB}$.)

\section{Proof of \eqref{eq:Z2_Y2_k}}
\label{sec:Z2_Y2_k_proof}

We start with the term $\tfrac{1}{2} \Tr[W^{k}_{\rB} Y_{\rB}]$. In an
appropriate phase convention, the operator $Y$ can be expressed as
\begin{equation}
  Y = \ee^{\ii \tfrac{\varphi}{2}} \trans{\rz}{\rz'}
  + \ee^{-\ii \tfrac{\varphi}{2}} \trans{\rz'}{\rz} \,,
\end{equation}
where $\ket{\rz}$ and $\ket{\rz'}$ are the eigenstates of $Z$ such that
$Z = \proj{\rz} - \proj{\rz'}$. In terms of these $\rz$ states,
\begin{IEEEeqnarray}{rCl}
  \label{eq:trWY_abszz}
  \tfrac{1}{2} \Tr[W^{k}_{\rB} Y_{\rB}]
  &=& \tfrac{1}{2} \Tr \bigsq{(W^{k}_{\rB} \otimes \id_{\rE}) Y}
  \IEEEnonumber \\
  &=& \re \bigsq{\ee^{-\ii \tfrac{\varphi}{2}} \bra{\rz} W^{k}_{\rB}
    \otimes \id_{\rE} \ket{\rz'}} \IEEEnonumber \\
  &\leq& \babs{\bra{\rz} W^{k}_{\rB}
    \otimes \id_{\rE} \ket{\rz'}} \,.
\end{IEEEeqnarray}
We let $U_{\rE}$ be any Hermitian unitary operator acting on
$\hilb_{\rE}$. Such an operator can always be expressed as the difference
between two orthogonal projectors, which we call $P_{\rE}$ and $Q_{\rE}$,
such that $U_{\rE} = P_{\rE} - Q_{\rE}$. Inserting
$\id_{\rE} = P_{\rE} + Q_{\rE}$ into the last line of \eqref{eq:trWY_abszz}
and developing, we obtain
\begin{IEEEeqnarray}{rCl}
  \tfrac{1}{2} \Tr[W^{k}_{\rB} Y_{\rB}]
  &\leq& \babs{\bra{\rz} W^{k}_{\rB} \otimes P_{\rE} \ket{\rz'}}
  + \babs{\bra{\rz} W^{k}_{\rB} \otimes Q_{\rE} \ket{\rz'}} \IEEEnonumber \\
  &\leq& \sqrt{\bra{\rz} \id^{k}_{\rB} \otimes P_{\rE} \ket{\rz}}
  \sqrt{\bra{\rz'} \id^{k}_{\rB}
    \otimes P_{\rE} \ket{\rz'}} \IEEEnonumber \\
  &&+\> \sqrt{\bra{\rz} \id^{k}_{\rB} \otimes Q_{\rE} \ket{\rz}}
  \sqrt{\bra{\rz'} \id^{k}_{\rB}
    \otimes Q_{\rE} \ket{\rz'}} \IEEEnonumber \\
  &\leq& \sqrt{\bra{\rz} \id^{k}_{\rB} \otimes P_{\rE} \ket{\rz}
    + \bra{\rz'} \id^{k}_{\rB} \otimes Q_{\rE} \ket{\rz'}} \IEEEnonumber \\
  &&\times\> \sqrt{\bra{\rz'} \id^{k}_{\rB} \otimes P_{\rE} \ket{\rz'}
    + \bra{\rz} \id^{k}_{\rB} \otimes Q_{\rE} \ket{\rz}} \,,
  \IEEEeqnarraynumspace
\end{IEEEeqnarray}
in which we used the Cauchy-Schwarz inequality and that
$(W^{k}_{\rB})^{2} = \id^{k}_{\rB}$ to obtain the second and third
lines. Substituting $P_{\rE} = \tfrac{1}{2} (\id_{\rE} + U_{\rE})$ and
$Q_{\rE} = \tfrac{1}{2} (\id_{\rE} - U_{\rE})$,
\begin{IEEEeqnarray}{rCl}
  \tfrac{1}{2} \Tr[W^{k}_{\rB} Y_{\rB}]
  &\leq& \sqrt{p_{k}
    + \tfrac{1}{2} \Tr \bigsq{(\id^{k}_{\rB} \otimes U_{\rE}) Z}}
  \IEEEnonumber \\
  &&\times\> \sqrt{p_{k}
    - \tfrac{1}{2} \Tr \bigsq{(\id^{k}_{\rB} \otimes U_{\rE}) Z}}
  \IEEEnonumber \\
  &=& \sqrt{p\subsup{k}{2} - \tfrac{1}{4} \Tr \bigsq{
      (\id^{k}_{\rB} \otimes U_{\rE}) Z}^{2}} \,, \IEEEeqnarraynumspace
\end{IEEEeqnarray}
where we recovered $\mathcal{I} = \proj{\rz} + \proj{\rz'}$, $Z = \proj{\rz}
- \proj{\rz'}$, and $p_{k} = \tfrac{1}{2} \Tr[\id^{k}_{\rB}
\mathcal{I}_{\rB}]$. The end result rearranges to \eqref{eq:Z2_Y2_k} in the
main text.

\section{Tightness of the min-entropy bound}
\label{sec:bound_tightness}

The lower bound \eqref{eq:hmin_chsh} on the conditional min entropy is tight
and is attained with a PM version of the optimal collective attack given in
\cite{ref:ab2007}, which we describe here. We set the states $\rho_{x,a}$
(following Eve's attack) to $\rho=\ket{\alpha}\bra{\alpha}$ and
$\rho'=\ket{\alpha'}\bra{\alpha'}$, where
\begin{IEEEeqnarray}{rCl+rCl}
  \ket{\alpha} &=& \ket{0}_{\rB} \ket{\psi}_{\rE} \,, &
  \ket{\alpha'} &=& \ket{1}_{\rB} \ket{\psi'}_{\rE} \,,
\end{IEEEeqnarray}
in which $\ket{0}_{\rB}$ and $\ket{1}_{\rB}$ are orthonormal and
$\ket{\psi}_{\rE}$ and $\ket{\psi'}_{\rE}$ are normalised states whose inner
product defines the specific attack. We set $\braket{\psi}{\psi'} = F_{\rz}$
for some real constant $0 \leq F_{\rz} \leq 1$. We also set
$\sigma=\ket{\beta}\bra{\beta}$ and $\sigma'=\ket{\beta'}\bra{\beta'}$ with
\begin{IEEEeqnarray}{rCl}
  \ket{\beta} &=& \frac{1}{\sqrt{2}}
  \bigro{\ket{\alpha} + \ket{\alpha'}} \,, \\
  \ket{\beta'} &=& \frac{1}{\sqrt{2}}
  \bigro{\ket{\alpha} - \ket{\alpha'}} \,.
\end{IEEEeqnarray}
With these definitions, the source states span a qubit subspace. Note that
$\braket{\alpha}{\alpha'} = \braket{\beta}{\beta'} = 0$, such that
$\alpha = \beta = 1$.

From the above definitions, we have that $\rho\0_{\rE} = \proj{\psi}_{\rE}$
and $\rho\1_{\rE} = \proj{\psi'}_{\rE}$, and thus
\begin{equation}
  \label{eq:trdist_Fz}
  D(\rho\0_{\rE}, \rho\1_{\rE}) = \sqrt{1 - F\subsup{\rz}{2}} \,.
\end{equation}
For the operators $Z = \proj{\alpha} - \proj{\alpha'}$ and $X = \proj{\beta}
- \proj{\beta'} = \trans{\alpha}{\alpha'} + \trans{\alpha'}{\alpha}$, we find
the partial traces
\begin{equation}
  Z_{\rB} = \proj{0}_{\rB} - \proj{1}_{\rB} = \sz
\end{equation}
and
\begin{equation}
  X_{\rB} = F_{\rz} \bigro{\trans{0}{1}_{\rB} + \trans{1}{0}_{\rB}}
  = F_{\rz} \sx\,.
\end{equation}
For optimal measurements on Bob's side, we can write
\eqref{eq:chsh_cc_trexpr} as
\begin{IEEEeqnarray}{rCl}
  S &=& \tfrac{1}{2} \trnorm{Z_{\rB} + X_{\rB}}
  + \tfrac{1}{2} \trnorm{Z_{\rB} - X_{\rB}} \IEEEnonumber \\
  &=& \tfrac{1}{2} \trnorm{\sz + F_{\rz} \sx}
  + \tfrac{1}{2} \trnorm{\sz - F_{\rz} \sx} \IEEEnonumber \\
  &=& 2 \sqrt{1 + F\subsup{\rz}{2}} \,,
\end{IEEEeqnarray}
which rearranges to
\begin{equation}
  F_{\rz} = \sqrt{S^{2} / 4 - 1} \,.
\end{equation}
Combining with \eqref{eq:trdist_Fz} confirms that we have described a family
of attacks for which $D(\rho\0_{\rE}, \rho\1_{\rE}) = \sqrt{2 - S^{2} / 4}$.

\bibliography{semidev}

\end{document}